\documentclass[review]{elsarticle}

\usepackage{lineno}
\usepackage{caption}
\usepackage{subcaption}
\usepackage{amsmath}
\modulolinenumbers[5]
\usepackage{graphicx}

\usepackage[%
    colorlinks=true,
    linkcolor=red
]{hyperref}

\journal{Journal of \LaTeX\ Templates}

\begin{document}

\begin{frontmatter}

\title{Improving the light collection using a new NaI(Tl) crystal encapsulation.}

\author[a]{J.J.~Choi}
\author[b,d]{B.J.~Park}
\author[c]{C.~Ha}
\cortext[mycorrespondingauthor]{corresponding author}
\ead{chha@cau.ac.kr}
\author[d]{K.W.~Kim}
\author[a]{S.K.~Kim}
\author[d,b]{Y.D.~Kim}
\author[d]{Y.J.~Ko}
\author[d,b]{H.S.~Lee}
\author[b,d]{S.H.~Lee}
\author[d]{S.L.~Olsen}

\address[a]{Department of Physics and Astronomy, Seoul National University,\\Seoul 08826, Republic of Korea}
\address[b]{IBS School, University of Science and Technology (UST), \\Daejeon 34113, Republic of Korea}
\address[c]{Department of Physics, Chung-Ang University,\\ Seoul 06973, Republic of Korea}
\address[d]{Center for Underground Physics, Institute for Basic Science (IBS),\\Daejeon 34126, Republic of Korea}

\begin{abstract}
  NaI(Tl) crystals are used as particle detectors in a variety of rare-event search experiments
  because of their superb light-emission quality.
  The crystal light yield is generally high, above 10 photoelectrons per keV,
  and its emission spectrum is peaked around 400~nm, which matches well to the sensitive region of
  bialkali photocathode photomultiplier tubes.
  However, since NaI(Tl) crystals are hygroscopic, a sophisticated method of encapsulation has to be applied
  that prevents moisture from chemically attacking the crystal and thereby degrading the emission.
  In addition, operation with low energy thresholds, which is essential for a number of new phenomenon searches,
  is usually limited by the crystal light yield; in these cases higher light yields can translate into
  lower thresholds that improve the experimental sensitivity.
  Here we describe the development of an encapsulation technique that simplifies the overall design by
  attaching the photo sensors directly to the crystal so that light losses are minimized.
  The light yield of a NaI(Tl) crystal encapsulated with this technique was improved by more than 30\%,
  and as many as 22 photoelectrons per keV have been measured.
  Consequently, the energy threshold can be lowered and the energy resolution improved.
  Detectors with this higher light yield are sensitive to events with sub-keV energies and well suited for
  low-mass dark matter particle searches and measurements of neutrino-nucleus coherent scattering.
\end{abstract}


\end{frontmatter}

\section{Introduction}
It is well established that dark matter exists in our universe and corresponds to 26.4\% of its total energy
content~\cite{PhysRevD.98.030001}.
Weakly Interacting Massive Particles (WIMPs) are dark matter candidate particles that are frequently considered
because of the balance among their inferred relic abundance, the measured dark matter density and the strength
of the weak interactions~\cite{PhysRevLett.39.165,Jungman:1995df}. 
The scattering of relic WIMPs in the galactic dark matter halo from ordinary nuclei is being searched for
in a number of experiments with a variety of target nuclei.

The experimental signature for these searches is
the detection of the recoil nucleus. For WIMPs with masses in the range between a few hundred MeV and a few
hundred GeV, the recoil nuclei have kinetic energies of a few keV and the current state of the art searches 
have set upper limits on the WIMP-nuclei interaction cross-sections of $\rm 10^{-46}~cm^2$ to
$\rm 10^{-40}~cm^2$~\cite{Schumann:2019eaa}. Recoil energies of low-mass WIMP nuclei scattering would typically
be in the sub-keV energy range, and ultra-low-threshold experiments are needed to probe these masses.
Interestingly, low-energy neutrino-nuclei scattering has the
same experimental signature as WIMP-nuclei interactions, where the neutrinos might originate from man-made
accelerators, nuclear reactors, or the cosmos.  An incoming neutrino with an energy of a few hundred MeV
or less can interact coherently with the entire target nucleus with a cross section that can be as high as
$\rm 10^{-40}~cm^2$.  The first unambiguous detection of the coherent neutrino-nuclei scattering was reported in
2017~\cite{Akimov1123}, some 40 years after it was first predicted~\cite{PhysRevD.9.1389}.
The count rates of these rare signals decrease exponentially with increasing nuclear recoil energy.
Thus, experiments with low-thresholds and high-light yields can, in general, improve the sensitivity of
both low-mass WIMP searches and coherent neutrino scattering measurements.

Thallium-doped sodium iodide crystals (NaI(Tl))~\cite{PhysRev.74.100} are suitable for low-energy-threshold
rare-event search experiments.  In the currently operating COSINE-100 WIMP-search experiment, NaI(Tl) crystals
with light yields of 15 PEs/keV~\footnote{PE stands for photoelectrons.}
are operated with an energy threshold of 1~keV~\cite{Adhikari:2017esn,Adhikari:2018ljm,Adhikari:2020xxj}.
At that threshold, the main limitation is the low number of detected photo-electrons in the detected signal
pulses, which makes them difficult to distinguish from the copious photomultiplier tube (PMT) noise-induced
pulses. Therefore, to access events with energies below 1~keV, improvements in the light yield are needed.

It may be possible to improve the intrinsic light output of the crystal material itself by changes in the
crystal production process~\cite{Park:2020fsq}. This is a long-term program that addresses a number of
non-trivial technical issues that are currently being pursued in parallel with the detector encapsulation studies.
The encapsulation R\&D uses existing NaI(Tl) crystals and is being reported here.  Especially, we focus on improvements in the
efficiency for collecting of the radiation-generated scintillation photons by means of a simplified light coupling
scheme. Also, because of the hygroscopic properties of NaI(Tl) crystals, a small amount of moisture contaminations during
the detector assembly procedures can also affect the light yield. We, therefore, have given special attention
to the development of techniques that limit the crystal's exposure to humidity during the assembly procedure.

\section{Method}
\begin{figure*}[!htb]
  \begin{center}
    \includegraphics[width=0.95\textwidth]{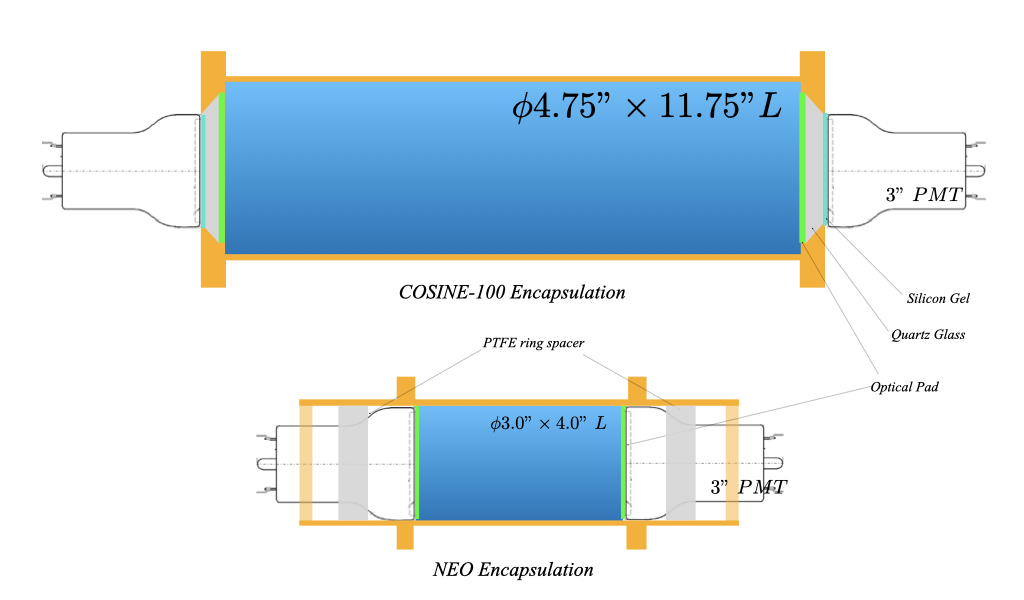}
  \end{center}
  \caption{The COSINE-100 detector design (top) 
    is 4.75-inches in diameter and is independently encased by copper with
    an optical pad and a quartz window inside of each end of the cylindrical module. The
    detector module is, in turn, coupled at each end to a 3-inch PMT via an optical gel.
    The new design (bottom) encapsulates a 3-inch diameter crystal and a 3-inch PMT that is
    directly coupled to the NaI(Tl) end-surface via a single optical pad.
    There is a PTFE ring spacer with 1~cm thickness (grey) that pushes against the PMT to give pressure on the optical pad. 
  }
  \label{newdesign}
\end{figure*}

The new encapsulation design principle that we follow is the minimization of light losses during the transit of
scintillation photons from their generation site to the PMT photocathode.
For comparison, we use the existing COSINE-100 crystals which are cylindrical with dimensions of  4.75-inches in diameter and
11.75-inches in length. These crystals have a rough-polished, reflector-wrapped lateral area and clear polished end faces. 
An optical pad made of silicone elastomer and a quartz glass window are coupled to each end of the crystal
and the crystal assembly is encapsulated within an airtight copper housing (see Fig.~\ref{newdesign}).
A 3-inch~PMT is attached to the quartz window via an optical gel.

In the COSINE-100 detector modules, shown in the top panel of Fig.~\ref{newdesign}, generated photons
that are incident on the portion of the outer edge of the 4-inch quartz window in 4.75-inch crystal that is
not covered by the 3-inch PMT photocathode have a low probability of being detected. By matching the size
of the crystal end face to that of the PMT photocathode, all of the generated photons that are incident
on the endface of the crystal are collected with high efficiency. In addition, we carefully polish the
entire crystal surface and use only a single optical pad between the PMT window and the NaI(Tl) end surface.
This reduces light losses due to reflections at each optical surface. The design of this new encapsulation
configuration is shown in the bottom panel of Fig.~\ref{newdesign}.

The COSINE-100 detectors are a detector-sensor separated design while
the new design is a detector-sensor combined assembly where the PMTs are
integral components of the airtight crystal encapsulation system.
This removes the quartz window and the optical gel but at the cost
of a more difficult encapsulation procedure. The NaI(Tl) crystal's vulnerability to moisture requires
a tight seal that is secure from any air leakage, while applying a limited amount of pressure onto
the relatively fragile PMT structure. To accomplish this, we placed a 1~cm-thick PTFE ring spacers
shaped to fit the neck part of PMT glass between the endcap
of the copper cylinder and the back of the PMT glass envelope that applied just enough pressure to couple
the optical pad to the crystal end face and maintain the airtight integrity of the encapsulated structure.
The force is applied via four screws to the back of the PTFE using a preset torque wrench.
Since any leaked air would quickly degrade the crystal's surface quality and, thereby, reduce the light
output, we use the measured light yield as the primary monitor of the long-term stability and
airtightness of the assembly.

We have tested three crystals with these encapsulations that are labeled as NEO-1, NEO-2, and NEO-3.

\subsection{Crystal size matching with the PMT photocathode}
\begin{figure*}[!htb]
  \begin{center}
    \includegraphics[width=0.9\textwidth]{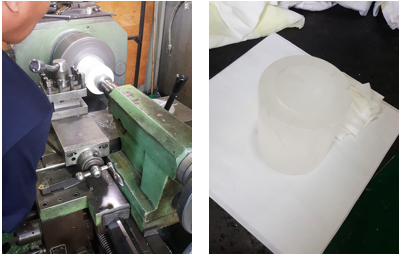}
  \end{center}
  \caption{NEO-1 crystal machining.
    Rough machining was used to reduce the diameter of a 4-inch crystal to 3-inches in a
    normal atmosphere with a lathe.
    While machining, we had to pay special attention because the brittleness of the crystal
    and weakness for its limited ability to support stress. The crystal was turned with a very
    sharp tool bit at a slow revolution speed.
  }
  \label{machining}
\end{figure*}
In this development, we use 3-inch low-background PMTs (R12669SEL) that have a high (40\%) quantum
efficiency for 400 nm photons. The crystal components were cut to match the 3-inch
photocathode.  For the first crystal, NEO-1, the original size of the crystal ingot was
4-inches in diameter; its diameter was reduced to 3-inches using a lathe, as shown in Fig.~\ref{machining}.
On the other hand, the 4-inch length of the crystal is chosen to keep as much of the original ingot length as possible.
In COSINE-100, we have seen no noticeable effect in light yield due to the length of crystal when two crystals with twice different lengths are compared~\cite{Adhikari:2017esn}. 
Since the NaI(Tl) crystal is brittle, we had to pay special precautions during the machining,
but some cracks were inevitable in our first attempt.
After seeing the results of NEO-1, for the next two crystals, NEO-2 and NEO-3,
we designed and ordered 3-inch diameter cylinder detectors and
measured the original light yields prior to re-encapsulation.

\subsection{Crystal polishing}
\begin{figure*}[!htb]
  \begin{center}
    \includegraphics[width=0.45\textwidth]{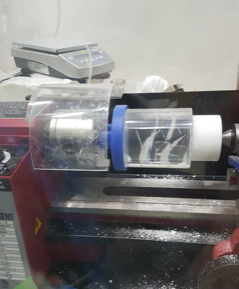}
    \includegraphics[width=0.45\textwidth]{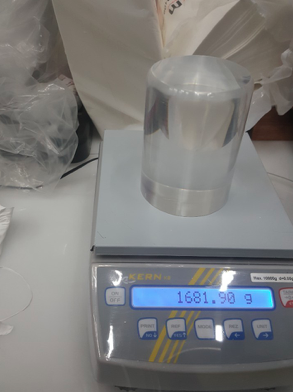}
  \end{center}
  \caption{Fine polishing inside the low-humidity glovebox (left) and the final product (right).
    For the polishing, a small lathe with lapping papers was used.
    All surfaces were polished until they were of optical quality.
  }
  \label{polishing}
\end{figure*}

The next step was to polish all of the surface areas of the 3-inch crystals. The lateral
areas are included in this procedure because we found that the roughened surfaces of the COSINE-100
crystals had some radioactive surface contamination that originated from either the polishing film or the
environment~\cite{Park:2020fsq}.  We did the polishing in a low-humidity glovebox that was continuously flushed with
nitrogen gas with the humidity maintained below 100 ppm$_v$ of H$_2$O by means of a molecular sieve
trap.   The polishing was done in several stages with lapping papers of different grits
using a small lathe that was located inside the glovebox (see Fig.~\ref{polishing}).
As soon as the crystal polishing was finished, the encapsulation procedure was started in order
to minimize additional radioactive contaminants on the surfaces of the crystal.
Also, we used carefully cleaned and dried encapsulation  components that had been baked at high
temperature and kept in the glovebox for a long enough time so that the level of H$_2$O emanation from them
was too low to affect the crystal's surface quality.

\subsection{Crystal light coupling}
As shown in the top diagram of Fig.~\ref{newdesign}, one end of the COSINE-100 crystal
is coupled through an optical pad, a quartz window, and an optical gel in series
to the PMT photocathode. The optical pad~\footnote{EJ-560 from Eljen Technology}
has 90\% light transmission for 400~nm photons and a refractive index of 1.43;
the optical gel~\footnote{EJ-550} is nearly transparent with a similar refractive index.
A typical quartz glass with 1~cm thickness has about 93\% transmission.
Therefore, in the old design, we expect that at least 15\% of generated photons
do not reach to the photocathode.
In the new design, by directly attaching the PMT to a crystal only using an 3-mm thick optical pad,
we eliminate the absorption in the quartz and minimize the loss of the photons due to reflection in the material interfaces.

\begin{figure*}[!htb]
  \begin{center}
    \includegraphics[width=0.9\textwidth]{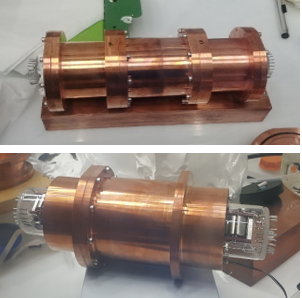}
  \end{center}
  \caption{The NEO-1 (top) and NEO-3 (bottom) detectors.
    In both detectors, the PMTs are sealed along with the polished crystal ingots.
    The crystals are in the middle section between the copper flanges that, together with
    the PMT glass envelope body, form the airtight seal.
  }
  \label{neo1_3}
\end{figure*}

The crystal detector assemblies are shown in Fig.~\ref{neo1_3}.

\section{Measurements}
To measure the crystal light yields and resolutions, we used a simple test setup in a surface-level
laboratory that included a 4$\pi$, 20~cm-thick lead and 5~cm-thick copper shield against environmental
background radiation.
A $^{241}$Am source located at the middle of the crystal scintillator provided
a 59.5~keV gamma line that is produced during its alpha transition to $^{237}$Np.

Additionally, we tested the detectors in a facility at the Yangyang Underground Laboratory~(Y2L),
where the cosmic-ray muon rate is strongly suppressed by the 700~m rock overburden, and
shielding comprised of lead, copper, and polyethylene attenuated the environmental radiation.
The Y2L setup has 12~low-background CsI(Tl) crystals that surround the test volume that are used
to tag accompanying radiation, which facilitates the evaluation of internal backgrounds in
the NaI(Tl) crystal that is being studied.

\begin{figure*}[!htb]
  \begin{center}
    \includegraphics[width=0.45\textwidth]{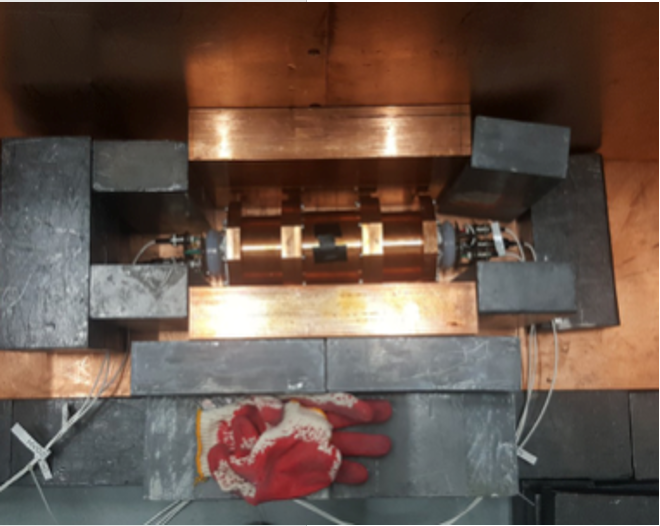}  
    \includegraphics[width=0.45\textwidth]{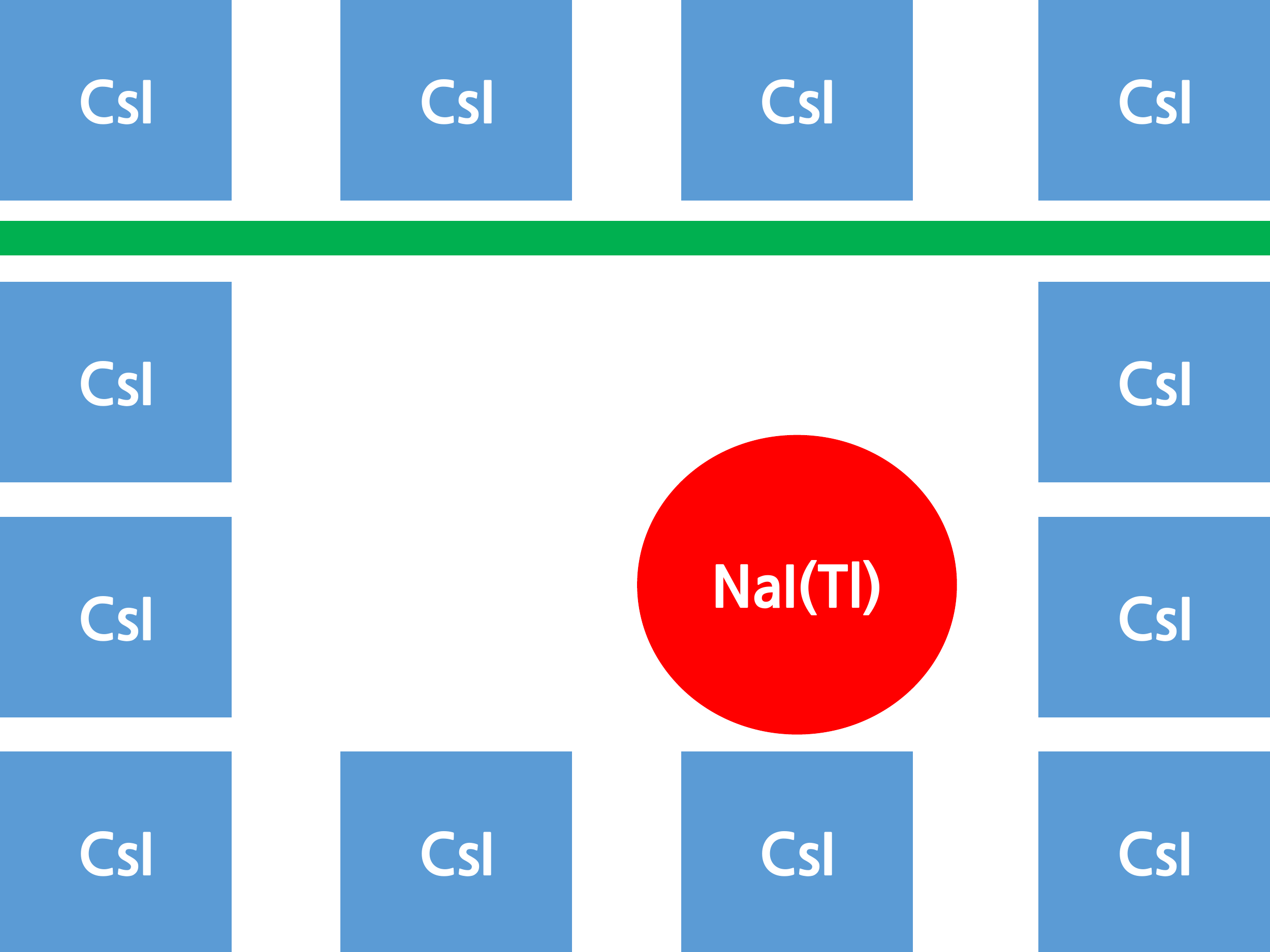}
  \end{center}
  \caption{A photo of the surface laboratory setup (left) and the schematic of the 700~m underground setup at Y2L (right).
    The Y2L setup contains rectangular-shaped CsI(Tl) crystals that facilitate the  identification of
    background contaminants in the NaI(Tl) crystal that is being tested.
    The green horizontal bar made of PMMA on the right schematic supports the top row CsI(Tl) crystal detectors.}
  \label{beforeanfter}
\end{figure*}

Figure~\ref{beforeanfter} shows the surface laboratory setup and the Y2L setup that are used in these tests.

\section{Results}
\subsection{Before and after the polishing and simplified coupling}
The shape of the waveform produced by single photoelectrons was characterized using isolated signals in
the tail part of NaI(Tl) pulses associated with 59.5~keV gamma rays for which the full energy is
deposited in the crystal.  From this, the light yield is determined from the ratio of the total deposited
charge to the single photoelectron's (SPE) mean charge, scaled to 59.5~keV.

NEO-1 was made from a 4-inch diameter crystal with a one-window encapsulation. So, for this
test crystal, the disentangling of the effects of crystal resizing and the modified light coupling
was difficult.
On the other hand, NEO-2 and NEO-3 started out  with 3-inch diameter ingots with a
one-window encapsulation made by the same vendor.  The measured light yields of the original detector
configurations were 10.7, 16.9, and 17.7~PEs/keV for NEO-1, NEO-2, and NEO-3, respectively.
The lower yield for the original NEO-1 measurement was likely
due to the size mismatch between the crystal end face and the PMT photocathode.

The light yields for these crystals after the re-encapsulation are measured to be 20.5, 19.3, and 21.8~PEs/keV.
For NEO-2 and NEO-3, the new design improves the light yields by 14\% and 23\%, respectively.
This improvement likely comes from the clear-polishing of the crystal combined with the simplified optical
coupling. It is likely that the marginal improvement for NEO-2 compared to NEO-3 is due to a few cracks developed
near the endface when the re-encapsulation was performed.
In case of the NEO-3 measurements, we have additionally verified the yields
with an SPE charge spectrum that was determined with a LED source.
The light yields are summarized in Table~\ref{table}.

\begin{table}[ht]
  \caption{The light yield measurements before and after the encapsulation change.
    The last column shows the light yield for one of the COSINE-100 crystals measured
    in the same way. NEO-1 shows a higher light yield after the resizing of the crystal.
    However, the original light yield of the 4-inch crystal was not
    accurately measured due to a mismatch between the crystal base size and PMT photocathode size.
    The units of the light yield measurements are PEs/keV. }
  \centering
  \begin{tabular}{c c c c}
    \hline
    NEO-1 & NEO-2          & NEO-3         & COSINE-100  \\
    after(before)&  after(before) & after(before) &   C6  \\
    \hline
    20.5$\pm$1.0 (10.7$\pm$0.7) &  19.3$\pm$0.9(16.9$\pm$0.9) &  21.8$\pm$0.9(17.7$\pm$0.9)  & 15.8$\pm$1.0\\
    \hline
        \label{table}
  \end{tabular}
\end{table}

\subsection{Comparison with the COSINE-100 crystals}
\begin{figure*}[!htb]
  \begin{center}
    \includegraphics[width=0.9\textwidth]{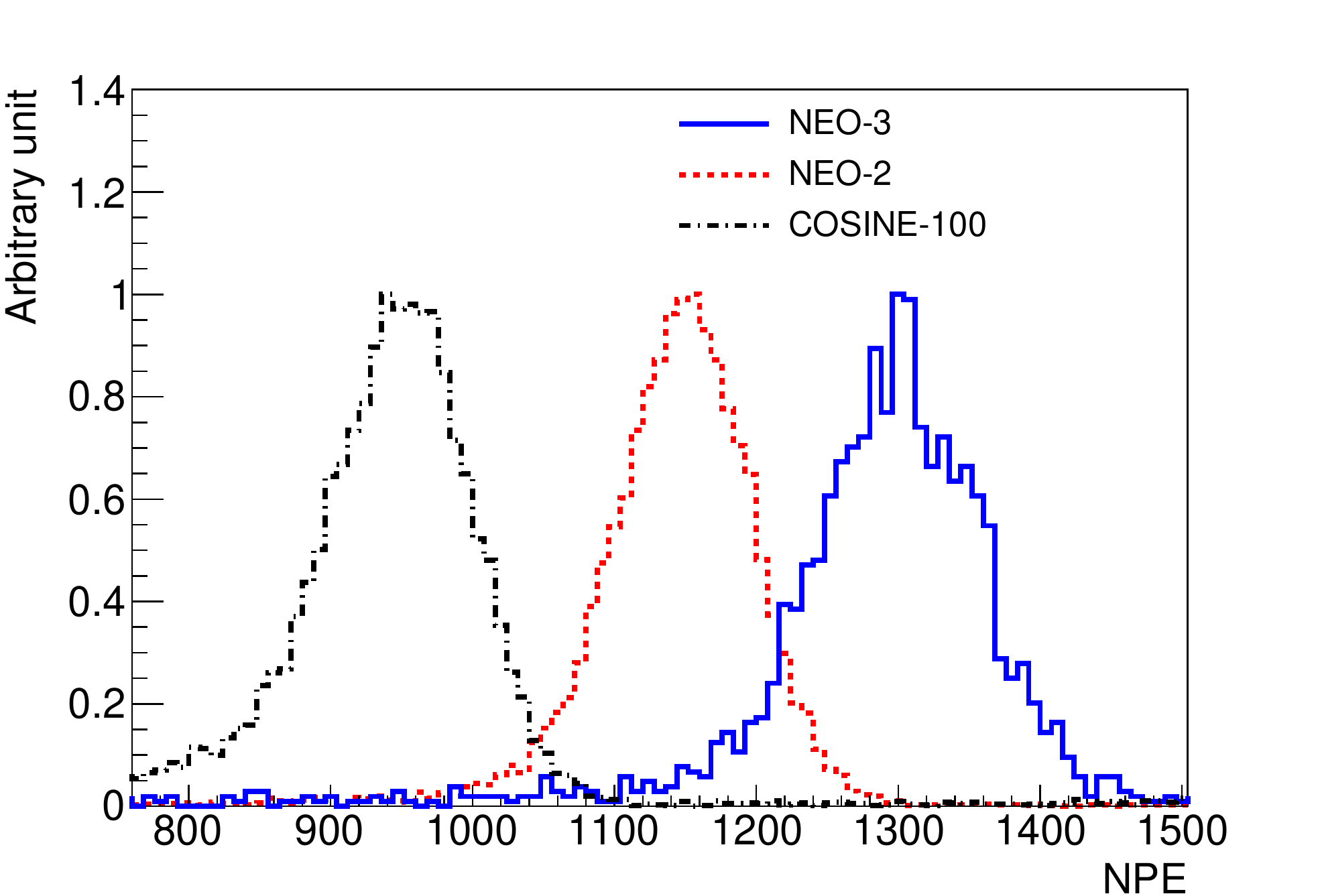}
  \end{center}
  \caption {The distributions of the number of photoelectrons (NPEs) associated the $^{241}$Am gamma peaks
    in the COSINE-100 (black dot-dashes), NEO-2 (red dashes) and NEO-3 (solid blue line) detectors. 
  } 
  \label{npe}
\end{figure*}

\begin{figure*}[!htb]
  \begin{center}
    \includegraphics[width=\textwidth]{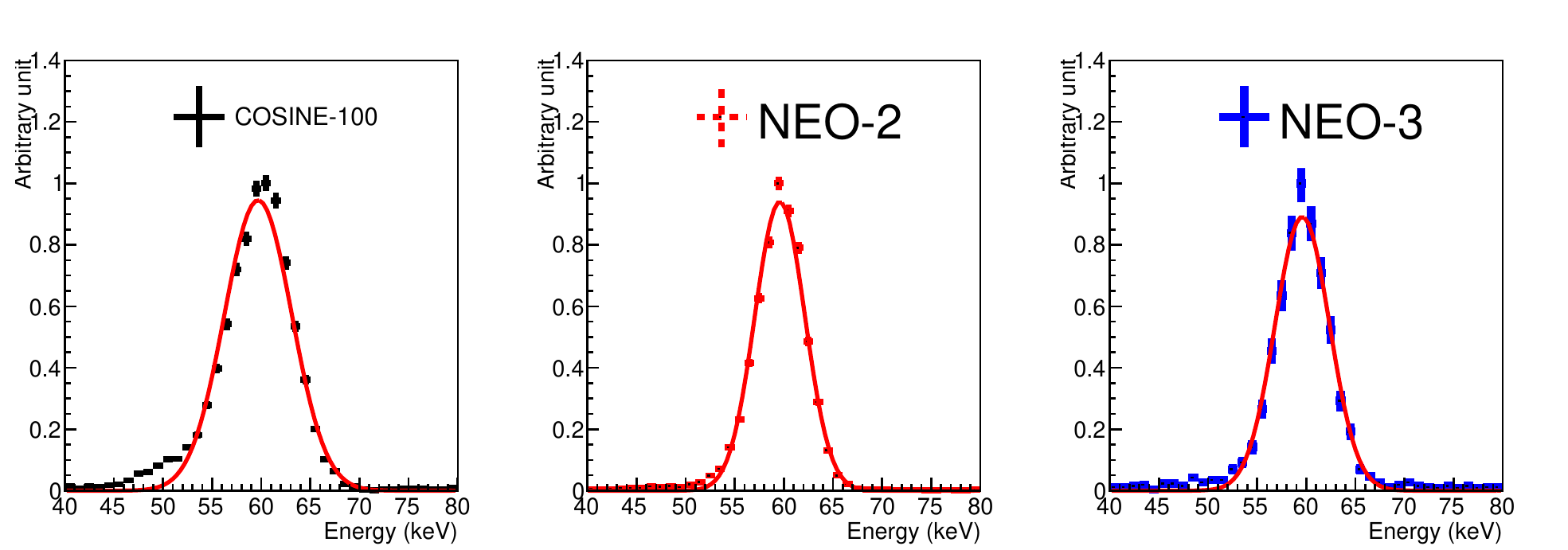}
  \end{center}
  \caption {The energy resolutions for 59.5~keV gamma rays of the three detectors.
    The black data points (left) are from COSINE measurements
    while the red (middle) from NEO-2 and the blue(right) from NEO-3, respectively.
    The resolutions are: 3.4~keV for the COSINE detector, 2.5~keV for NEO-2 and 2.8~keV for NEO-3
    obtained by Gaussian fits (red lines).
  } 
  \label{resolution}
\end{figure*}

Since the light yield directly affects the energy resolution of a crystal detector,
we compare the light yield and resolution of the peak 
with those from previous COSINE-100 measurements.  
Figure~\ref{npe} shows the light yield comparison between a COSINE crystal and the newly designed detectors.
Figure~\ref{resolution} shows the same peaks after the light-yield-to-energy calibration was applied. 
The new detectors have 30\%(NEO-1), 22\%(NEO-2), and 38\%(NEO-3) higher light outputs compared to the COSINE-100
crystal that has a light yield that was measured to be 15.8 PEs/keV using the same $^{241}$Am gamma peak, with
an energy resolution of $3.43\pm0.03(stat.)$~keV from Gaussian fits to the peak.
The NEO-2 and NEO-3 resolutions are determined in the same way
in Fig.~\ref{resolution} to be $2.54\pm0.02(stat.)$ and $2.78\pm0.06(stat.)$~keV, respectively.

\subsection{Long-term stability}
\begin{figure}
  \centering
  \includegraphics[width=\textwidth]{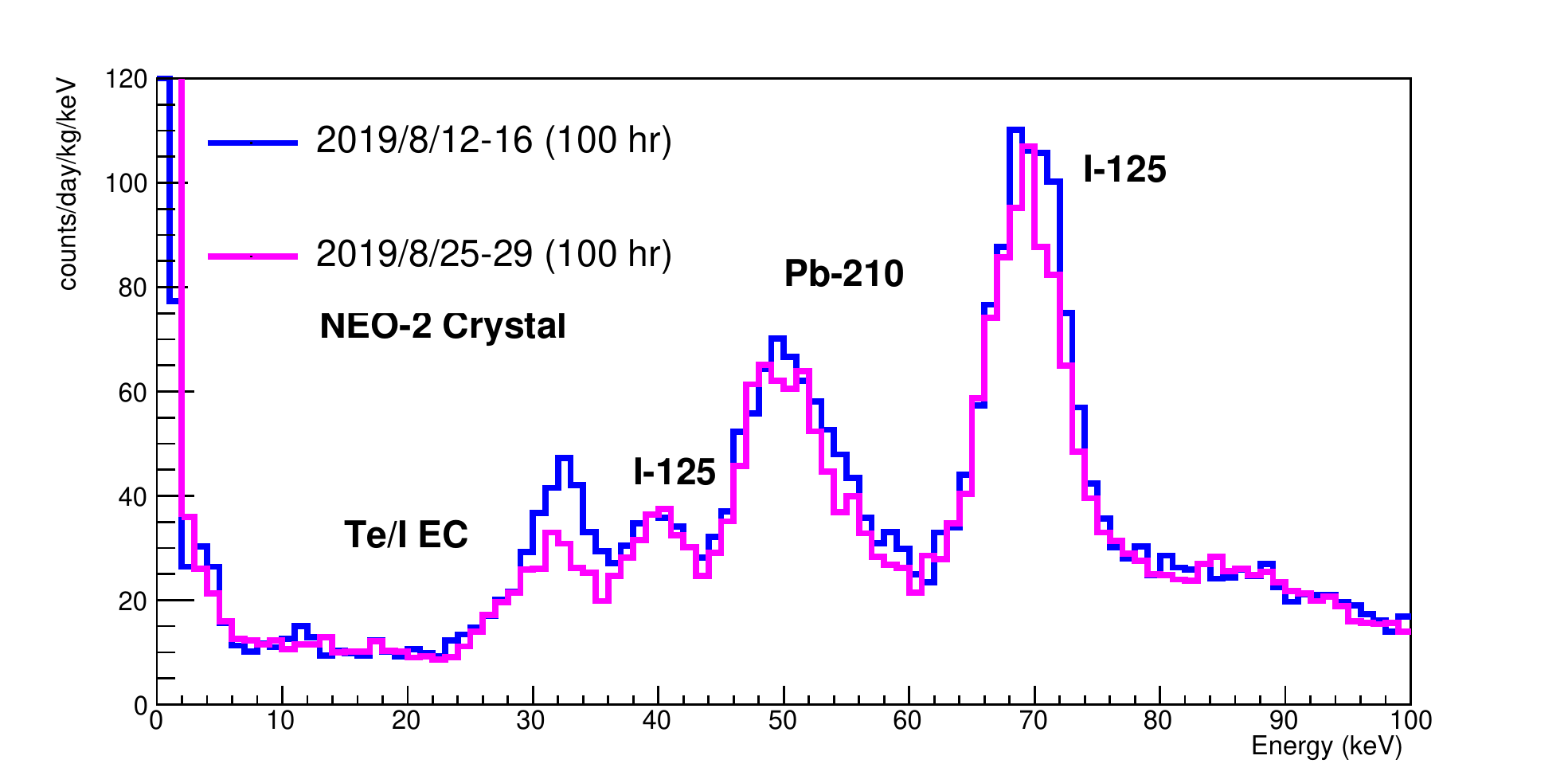}
  \caption{The low energy background spectra between 0 keV and 100 keV for two different time periods
    separated by two weeks. First 100 hours of data(blue) and the next 100 hours of data (magenta) are compared in cases of
    beta/gamma events. Several radioisotopes are decaying away as the cosmic activation is terminated
    in the underground laboratory. We apply the same energy calibration for the two data periods. Below 5 keV, there remained a residual noise contamination that obscured
    the beta/gamma spectrum. A modest event selection was applied to reject noise events and multiple-site events are
    removed using the surrounding CsI(Tl) veto detectors.
  }
  \label{stable}
\end{figure}

We have measured the NEO-2 energy spectrum at Y2L for a four week continuous period as a check on its stability.
For this we used crystal's internal peaks from $^{210}$Pb (46.5 keV gamma plus X-rays) and
cosmogenic $^{125}$I (67.2 keV) and $^{121m}$Te (30.5 keV)~\cite{deSouza:2019hpk}
to monitor the low energy spectrum.
Figure~\ref{stable} shows that the peak position did not change between first 100 hour data period and the succeeding 100 hour data
period, which indicates that the encapsulation does not have an air leak.

\section{Conclusion}
We have developed a method for NaI(Tl) crystal encapsulations that includes a well matched crystal-PMT window
size with a simplified light coupling design.
The results show  22--38\% light yield improvements and as much as 30\% improvement in energy resolution.
The absolute 22 PE/keV value in NEO-3 is by far the highest ever reported for a large-size NaI(Tl) crystal.
In addition, this new design performed stably during long-term stability checks.
We expect to use this technique for the fabrication of detectors for the next-generation, COSINE--200 phase of the experiment.
Studies are also underway of the feasibility of using them for reactor-based neutrino coherent scattering measurements with
a sub-keV energy threshold.

\section{Acknowledgments}
This research was funded by the Institute for Basic Science (Korea) under project code IBS-R016-A1; 
This research was supported by the Chung-Ang University Research Grants in 2020.

\end{document}